# An approach to systemic risks of AI through the lens of emergence, collective action problems, and externalities

Carsten Orwat,[1] Lucas Staab, Alexandros Gazos

Karlsruhe Institute of Technology,
Institute for Technology Assessment and Systems Analysis

20 Jul 2026

**Abstract:** The integration of general-purpose artificial intelligence models into downstream AI systems, among other developments, has given rise to new forms of risk that are more systemic in nature than conventional AI risks. However, there is no generally accepted definition of systemic risks in general and for AI in particular. Conceptualisations of these risks vary across research and regulation. Especially the application of the systemic risk approach to human rights or fundamental rights, like in the EU AI Act, is relatively new, just as the research on the contribution of AI to systemic forms of discrimination, privacy violations, erosions of democracy, or climate and environmental degradation. We argue that some concepts so far have not sufficiently take complexity and emergence into account. Furthermore, this variety of concepts might hinder responsible actors to adequately assess the systemic risks of AI, leading to inadequate prevention and mitigation measures and ineffective governance. To contribute to the understanding of systemic risks of AI, we propose a conceptualisation of systemic risks of AI that considers complex phenomena that lead to the emergence of harms at the societal or global level. We outline systemic risks mainly as complex externalities and collective action problems. Of particular interest are feedback dynamics, processes that lead to market concentration like network effects, algorithmic monocultures, and integration processes of AI supply chains or 'AI ecosystems' and across societal sectors, which can result in structural dominance, (inter-) dependencies, and cascading risks. Further phenomena contributing to systemic risks are information asymmetries, informational emergence, and deficits of the governance and institutional framework.

[1] Corresponding author (email address: orwat@kit.edu). Comments are welcome.

# 1 Introduction

The integration of general-purpose AI models into downstream AI systems, among other developments, has given rise to new forms of risks that are more systemic in nature than conventional AI risks. Although these risks gain traction in research and regulation, there are widely varying understandings of them. In view of a diversity of understandings of systemic risks in general and systemic risks of artificial intelligence (AI) in particular, we consider a proper conceptualisation of systemic risks of AI as essential because it influences conclusions about adequate approaches to assessment, prevention, mitigation, and resilience as well as governance requirements and options. In order to avoid neglecting or overlooking important causal patterns, manifestations, and harms of systemic AI risks, an appropriate conceptualisation should also pay attention to the 'systemic nature' of these risks.

They are systemic in nature, precisely because they threaten societal goals as forms of common goods, which are themselves emergent effects of complex interactions among a multitude of elements within a societal or global system. Previous research has developed similar perspectives in other fields of research dealing with comparable problems and challenges (see Section 2 and 3). Systemic risks are then failures of complex social and sociotechnical processes that would be necessary to provide such emergent common goods.

As the diverse understandings of systemic AI risks are nascent, the following sections mainly provide an informed conceptualisation and abstraction from common themes in previous research; this is neither a complete inventory of the range of systemic risks of AI nor a critical examination of inconsistencies of concepts, but rather an outline of common systemic risk phenomena and a proposal for an analytical heuristic to be further expanded upon. The paper applies a sociotechnical perspective considering not only technical systems but also social actors, their actions, the rules, routines, institutional and social structures, and their interrelations (e.g. Trist 1981; Selbst et al. 2019). This includes practices such as business models, governmental or administrative practices, social structures, and institutional and regulatory frameworks and processes, as well as the interactions between these elements as causes and pathways of systemic risks.

Since AI is a general-purpose technology with a multitude of different applications, it affects various societal or global systems, such as industries or sectors (e.g. the healthcare or financial systems), entire societies, various legal, political, or ethical systems (e.g. democratic systems), different jurisdictions, or the global Earth system. Furthermore, the paper considers various forms and developments of AI, including predictive AI, generative AI, general-purpose AI (GPAI) models, foundation models, agentic AI, and interactive agentic AI. In accordance with the protection goals of the European Union AI Act (Article 1, Regulation 2024/1689), we see fundamental rights, including environmental protection, democracy, and the rule of law, safety, and health, as the societal goals deemed valuable for the assessment, management, and governance of systemic risks.

The paper proceeds as follows: Section 2 outlines the different understandings of systemic risks. Section 3 introduces emergence as a feature of complex systems that is particularly relevant for the understanding of systemic risks of AI. Furthermore, our understanding relates emergence in sociotechnical systems to complex externalities and collective action problems as core components of the systemic risk concept, which is discussed in Section 4. We continue by outlining risk phenomena in complex sociotechnical systems that drive, influence and shape the emergence of systemic risks, which are explained in Section 5. These phenomena include informational emergence, proxy differentiation and bias, integration along the AI value chains, coordination and cascading failures, market concentration, algorithmic monoculture, and outcome homogenisation. Additional phenomena are feedback dynamics, rebound effects, information asymmetries, normative ambiguities, and deficits in the institutional and regulatory framework. Section 6 discusses how systemic risk phenomena interact, how they accumulate, and how non-linearities and path dependencies might emerge, and implications for assessing and addressing the acceptability of systemic risks. Section 7 introduces challenges and open questions in assessing, preventing, and mitigating systemic risks, while Section 8 draws some conclusions with regard to the main findings.

## 2 Different understandings of systemic risks

Currently, there is no generally accepted definition of systemic risks in research (Liu & Renn 2025). Nevertheless, certain characteristics have frequently been identified in previous research on systemic risks, in particular the complexity of the system(s) under consideration, the resulting emergence, aggregate, compound, or cascading risks, non-linear dynamics in system developments or causal relations, stabilising or destabilising feedbacks, self-organisation, or the existence of tipping points at which systemic changes can impair or disrupt the functions of a system (e.g. Renn et al. 2022; Kaufman & Scott 2003; Centeno et al. 2015; Helbing 2013; Schwarcz 2008; Lucas et al. 2018; Gambhir et al. 2025; Arnscheidt et al. 2025; Lawrence et al. 2024; Schweizer & Juhola 2024; Sillmann et al. 2022; James 2017). However, the approaches differ in terms of characteristics they consider, emphasise or relate to one another, in order to distinguish systemic risks from other types of risk. They also approach severity differently, depending on whether systemic risks merely restrict the realisation of societal goals, erode them or lead to an outright collapse of a system in question. Research has also started to address the systemic risks of AI specifically, but also with different understandings of systemic risks, ranging from individual AI providers causing systemic risks to focusing on complex interactions between actors (e.g. Svetlova 2022; Kaminski 2019; NIST 2023; Teo 2025; Darius et al. 2025; Schön et al. 2025; Kondor et al. 2024; Maham & Küspert 2023; Cecchetti et al. 2025; Hopkins et al. 2025b; Correia de Carvalho 2025; Hacker et al. 2025; Uuk et al. 2024; Loi et al. 2025; Bengio et al. 2025; Bengio et al. 2026; Palumbo 2026). Furthermore, related research on societal harms (e.g. Smuha 2021) also provides insights into systemic risks of AI.

In regulation, systemic risks of AI are addressed, in particular, in the EU AI Act and defined as "a risk that is specific to the high-impact capabilities of general-purpose AI models, having a significant impact on the Union market due to their reach, or due to actual or reasonably foreseeable negative effects on public health, safety, public security, fundamental rights, or the society as a whole, that can be propagated at scale across the value chain" (Article 3(65) AI Act). The definition emphasises the high-impact capabilities of general-purpose AI (GPAI) models and presupposes that these capabilities continue to grow primarily with increasing computing power and size of the GPAI model.[1] Above a certain level, GPAI capabilities would come with catastrophic potential, such as loss of control over the AI's self-development, misuse by malicious users for large-scale cyberattacks, harmful mass manipulation, or the production of weapons. It is problematic that the AI Act neglects other types of AI models and systems and considers only GPAI models, their high impact capabilities, and their providers as sources of systemic risk.

From the literature, the different understandings of systemic risks can be abstracted and divided into three idealised types of actor constellations and their responsibilities. First, individual actors – particularly the developers of GPAI models – are regarded as responsible for systemic risks. This is the main approach underpinning the AI Act. Additionally, this type includes scenarios of AI systems enabling individual adversarial actors to potentially cause system-wide harm, for instance, by developing chemical or biological weapons. Second, approaches from the literature on systemic risks emphasise interactions and interdependencies between system elements as is the case with systemic risks in the financial system or in relation to climate change. They often apply perspectives form complexity science, which includes the focus on cascading failures, feedback loops and the structures, relations and practices of society as sources of systemic risks. Third, analyses of systemic risks in the financial system in particular[2] have shown that governance frameworks, with their institutions and regulations, can contribute to the emergence of systemic risks, especially, because they shape the interactions and interdependencies between system elements. We regard these three idealised types as complementary and overlapping in practice. It should be noted that feedback dynamics, amplification mechanisms, cascading errors, etc. can occur in all three types, although the 'systemic nature' is more pronounced for the latter two types.

## 3 Systemic risks and emergence

To conceptualise systemic risks of AI, we analyse systemic risks mainly as a matter of complexity and emergence in societal or global systems (Page 2015; Renn 2020; Renn et al. 2022; Centeno et al. 2015; Johansen & Rausand 2014:277-278; Arnscheidt et al. 2025; Büscher 2011; Rhoen &

---

[1] The Code of Practice on GPAI models provides detailed information on this understanding (European Commission 2025).

[2] Research on systemic risks in finance demonstrates that the institutional framework can also be a source, or even the root cause, of systemic risks, including factors such as institutional complexity, inappropriate incentives inducing moral hazard or excessive risk-taking (Schwarcz 2008; Chambers et al. 2019; Cecchetti et al. 2025), or that institutional risk mitigation measures (Ruhl 2014) or the law is an additional factor for complexity inducing systemic risks (Castellano 2024). Research on business responsibilities to protect human rights also indicates that laws affecting business practices can be the root causes of human rights violations (Marks 2011; Kampourakis & Lane 2025).

Feng 2018; Svetlova 2022; Tsvetkova et al. 2024; Kolt et al. 2025; Darius et al. 2025; Khoo 2025). In complex systems, the emergent properties at the macro level cannot be reduced to the features of the individual parts at the micro level (e.g. Mitchell 2009). The properties and behaviour of entire sociotechnical systems are influenced but not determined by the properties of the system's parts, since the behaviour of the overall system arises from the relationships and interactions within and outside the system. In other words, examining the individual parts of a system does not explain the overall behaviour of the system. Emergent properties can only be understood by considering the interconnections and interactions between the system's parts (e.g. transactions, functional interdependencies, feedback dynamics, or behavioural reactions), their relationships, structure, and mode of organisation, as well as their behaviour and implications.

The elements of complex sociotechnical systems include human actors and organisations, technical components (e.g. hardware and software, datasets, data processing activities), practices, or institutional arrangements, such as laws, technical guidelines or standards, regulatory frameworks, or systems of moral or legal rules, customs and conventions. In these systems, emergent social phenomena can be events, institutions, structures, or values. Emergent properties result from complex combinations of social or sociotechnical processes, such as the overall employment level or the innovativeness of a country (Sawyer 2005; Mayntz 2008). Typical for emergence research is the consideration of the mutual influence between the micro and macro levels.[3] For instance, the collective or aggregate actions at the micro level also shape the structure and institutional arrangements of a societal or global system at the macro level, and the structure (e.g. network structure, central nodes, or dependencies) and institutional arrangements at the macro level enable, incentivise, influence, or constrains the actions and interactions at the micro level.

A conceptualisation that takes complexity and emergence into account must include and define systemic risks accordingly. From this perspective, systemic risks arise in complex systems (1) as the effects of interactions between system elements at the micro level that threaten the realisation of emergent phenomena at the macro level that a society considers desirable or beneficial, or (2) as emergent effects that a society considers undesirable or harmful. From the perspective taken in this paper, systemic risks arise from the emergent effects of collective action problems or from risk-generating interactions among the elements of a complex sociotechnical system (externalities), their dynamics, and their interactions with other systems. They are the outcomes of different types of interactions between humans, between humans and AI systems, or between AI systems themselves, which threaten emergent societal goals or cause emergent societal harm. In the following, the emergent phenomena that a society considers valuable can be the functions of a societal or global system (e.g. the stable provision of finance as a function of the financial system), but also the concomitant protection and promotion of societal goals such as fundamental rights, democracy, or the protection of the environment.

From this perspective, assessing systemic risks means reconstructing the factors most likely to interact in such a way that detrimental effects may arise at the macro level. It involves assessing societal processes, structures, and events leading to the emergence of systemic risks, as well as those that prevent the emergence of phenomena desired by a society. Systemic risks cannot be explained as the mere sum of individual risks; their emergence and severity can also result either from the interaction of multiple risks or from accumulated risks whose severity may also grow over-proportionally or non-linearly beyond the severity of the sum of the risks.

[3] For analytical purposes, meso levels with functions mediating micro-macro processes can also be introduced (e.g. Frischlich et al. 2025).

Research has developed concepts that view societal goals as emergent phenomena, thereby providing insights into the understanding of systemic risks. Research on systemic discrimination and structural or systemic injustices regards discrimination across society as an emergent phenomenon and as a result of complex interactions of social processes and structures (Council of Europe 2020; Reskin 2012; Rivault 2024; Bohren et al. 2025). This research analyses the complex, dynamic, and interconnected individual phenomena, collective logics, characteristics, and amplification mechanisms (e.g. Brown et al. 2025; McMillon 2024; Durlauf et al. 2025). It also focus on institutions such as the legal system or education, and the spread of discrimination from one sphere to others through interactions between institutions (e.g. Bohren et al. 2025). Overall, discrimination as an emergent phenomenon at the societal level cannot be explained by examining individual instances of discrimination or by the 'sum' of individual cases, but by analysing the interrelations and interplay between individual forms of discrimination and risks factors that cause, mitigate, perpetuate, or exacerbate discrimination. Similar research approaches can be found in the field of sustainable development, where sustainability is viewed as an emergent property at the systemic level (e.g. Bolte et al. 2022), in safety engineering, where safety and security is treated as an emergent phenomenon (Young & Leveson 2013; Yang et al. 2017; Lintern & Kugler 2023). Democracy can similarly be understood as emerging from an interplay of institutions, processes, practices, and norms that enable democratic participation, representation, accountability, and public discourse (Mansbridge et al. 2012), each potentially impacted by AI (Jungherr 2023; Kreps & Kriner 2023; Cupać & Sienknecht 2024; Summerfield et al. 2025).

Interactions between humans and AI systems are mainly enabled, steered, and governed by the structure of the sociotechnical system (e.g. network structure, central nodes, or dependencies) and the institutional and regulatory environment providing incentives or constraints. With the ongoing digitalisation of societies, the features and behaviour of sociotechnical systems are increasingly shaped also by the features and behaviour of digital technologies, applications, use practices, routines, as well as the features of digital 'ecosystems' or value chains. Thus, the factors of systemic risks of AI result from the patterns[4] of the structure, interactions, and dynamics of digital sociotechnical systems. These patterns can include collective action problems, coordination failures (such as breakdowns in cooperation, negotiation, competition, conflict, or dominance) and social dilemmas, aggregated, complex and systemic externalities, market concentration, network effects, path dependencies, lock-in situations, as well as other systemic properties and relationships such as feedback dynamics, dependencies, or couplings. The patterns can provide insights into how systemic risks arise from the actions and interactions of actors at the micro level and manifest themselves at the macro level. Not all of these patterns are newly introduced by AI, but they can change, intensify, or recombine with the development, deployment, and use of AI.

[4] The term 'patterns' refers to the study of social mechanisms as regularities in the behaviour and characteristics of agents and societal structures that lead to specific outcomes at the macro or societal level (e.g. Hedström & Ylikoski 2010). This research is also applied to systemic risks (see Büscher 2011).

# 4 Collective action problems and externalities in complex situations

## 4.1 Collective action problems and their complexity

Collective action problems include problems of public goods, common pool resources or club goods, the prisoner's dilemma, and free-rider problems (Olson 1965; Ostrom 1990; Kollock 1998; Cornes & Sandler 1996; Mayntz 2002). The basic assumption is that the self-interest of individual actors may lead them to behave in ways that do not contribute to the common good or lead to collectively suboptimal or disastrous outcomes. Different types of goods are usually distinguished according to whether exclusion from the benefits of the good is feasible or not, and whether its consumption is highly rivalrous or a joint use (Ostrom & Ostrom 1977; Cornes & Sandler 1996). Public goods are characterised by the high difficulty of excluding potential beneficiaries, and their use is non-rivalrous, meaning that consumption by one person does not diminish the amount available to others (Ostrom 2010).[5] Since beneficiaries cannot be excluded, public goods or common pool resources are characterised by social dilemmas in the form of free-rider problems, which result in public goods not being provided in a socially optimal manner or in common pool resources being exhausted. To solve social dilemmas, some kind of collective action is necessary, organised either by the involved actors themselves or by an external authority, which can be the state or other governance actors.

Several societal goals that are threatened by systemic risks are conceptualised as a type of common good or collective action problem at the societal or global level (Renn et al. 2022:1905; Schweizer & Juhola 2024:4), such as climate stability or the protection of the earth's resources (Hardin 1968; Ostrom 1990; Dietz et al. 2003), financial stability (Schwarcz 2008; Linarelli 2017; Ülgen 2021), cybersecurity (Kianpour et al. 2022), general level of AI safety (LaCroix & Mohseni 2022; Blomquist et al. 2025; Olugbade 2025; de Neufville & Baum 2021),[6] the general (or common) level of realisation of fundamental or human rights (Kaul & Mendoza 2003; Smuha 2021; Felice & Fuguitt 2021), the general level of data protection and privacy (Specht-Riemenschneider

[5] Other types of goods are common pool resources (high difficulty of exclusion combined with high rivalry; sometimes referred to as 'collective goods'), club (or toll) goods (low difficulty of exclusion combined with low rivalry), and private goods (low difficulty of exclusion combined with high rivalry) (Ostrom 2010:644-645). Ostrom used the term 'subtractability of use' instead of 'rivalry of consumption'.

[6] Laufer et al. (2024) model bargaining with the option of free-riding in fine-tunig among GPAI model provider and domain specialist adopting the model.

2023), or the common good character of privacy (Regan 1995; Regan 2002; Fairfield & Engel 2017).

Common goods, including public goods, common pool resources, and club goods, can change their characteristics over time by becoming rivalrous through overuse or excludable, in particular, through technological advances (Mayntz 2002). They can also be designated as common goods through political decisions (Kaul & Mendoza 2003). In democratic societies that uphold human rights, these are often realised by the state as public goods. This means that no individual should be excluded from enjoying the benefits of human rights and that the good is provided in such a way that there is no rivalry in its use; in other words, the realisation of rights for one person must not come at the expense of their realisation for others.

Designating fundamental rights as a public good and having them provided by private actors would give rise to free-rider problems, which is why they are generally provided by the state. If, for instance, fundamental rights were realised primarily through court enforcement, the litigation costs for those affected by violations of these rights could be very high for the individual, while the benefits of establishing law through court decisions may accrue to society as a whole. Many of those affected or potentially affected might therefore wait for others to fight for the right, thereby behaving like free riders.

With regard to fundamental rights, the common standards of their protection, realisation, enforcement, and stability emerging at the macro level can be characterised as a public good, described by expressions such as a society's 'general level of data protection' (Specht-Riemenschneider 2023) or 'general level of equality or non-discrimination'. These societal goals are the emergent properties that arise at the macro level as a result of the interactions of the elements of the social system.

Collective action problems in the provision of common goods can be distinguished into local, small-scale, and large-scale collective dimensions, which can be viewed on a continuum without clear boundaries (Jagers et al. 2020). Under certain conditions, collective action problems might be solved by self-regulation through mechanisms of trust, reciprocity, and reputation built up through repeated interactions, as well as through the possibility of mutual monitoring of actions. In contrast – and more likely in the case of systemic risks of AI – large-scale collective action problems are characterised by a large number of actors, making coordination and cooperation more difficult or potentially leading to agency problems through the involvement of representatives. Further factors increasing complexity are the spatial distance and time lags between causes and aggregate effects that can hinder collective actions. Other factors are the anonymity of actors with reduced face-to-face communication, the weakened sense of accountability among actors for small or infinitesimal contributions, in particular when others have little opportunity to observe individual actions, and the heterogeneity of actors. There can be an additional layer of complexity, if collective action problems are interconnected and mutual influences arise between them. The lack of knowledge about the extent of common goods or about the decisions or actions of other actors as well as emotional and psychological detachment are further factors (Jagers et al. 2020). These factors are critical for analysing systemic risks in order to assess whether small-scale risks or large-scale and complex systemic risks are to be expected, and whether a reduction in systemic risks can be achieved through self-governance and cooperation or whether an external authority might be required.

## 4.2 Externalities and their complexity

Externalities are present "whenever the behavior of a person affects the situation of other persons without the explicit agreement of that person or persons" (Buchanan 1971:7; overview in Rayamajhee & Paniagua 2026). Externalities, also known as external effects, spillovers, or social costs, can be positive if a causative actor does not receive the full benefit that others enjoy, or negative if a causative actor does not bear the full costs that are imposed on others, including harms that cannot be expressed in monetary terms. Potential negative externalities refer to situations in which the beneficiaries of a risky action and the risk bearers diverge, as the former shift risks onto the latter (risk spillover) without the latter adequately sharing in the benefits, being compensated for bearing the risk, or being able to consent to it. Besides shifting risks onto other individuals or groups, actors can also shift risks in time (e.g. by leaving the resolution of problems to future generations) or to other domains (e.g. by shifting infrastructure risks to sectors that depend on the infrastructure), thereby generating potential externalities.

The concept of systemic externalities (Wagner 2010) refers to externalities that arise from interactions between actors, e.g. through failure spillovers from one actor to another or when two or more actors fail to coordinate common risk mitigation and thereby generate externalities. They are also characterised by their impact at the macro level. Much like systemic externalities, complex and large-scale externalities are characterised by a large number of actors organised across multiple, interconnected, and nested levels, where it is difficult to identify causes and solutions (Paniagua et al. 2024; Paniagua & Rayamajhee 2024:396). Complex externalities, like pandemics, climate change, ocean pollution, or microplastic in ecosystems, can be viewed as emergent phenomena (Trantidis 2024), since a "complex externality involves a synergy of multiple and continually occurring activities producing a sizeable outcome whose effect, scale and severity cannot be simply disaggregated into each specific additive contribution" (Trantidis 2024:461).

Complex externalities also include 'nested externalities', which "occur when actions taken within one decision-making unit simultaneously generate costs or benefits for other units organized at different scales" (Ostrom 2012:356). In contrast to a dichotomous market-or-state approach to addressing externalities, mitigating the problems of complex externalities typically requires the efforts of many actors organised across multiple, interconnected, or nested levels, an approach often referred to as polycentric governance (Ostrom 2010; Paniagua et al. 2024; Paniagua & Rayamajhee 2024; Helberger 2026).

Although externalities and common goods are analytically distinct, negative externalities can hinder or destroy the provision or preservation of common goods, as e.g. in the case of greenhouse gas emissions, which cumulatively deplete the common good of climate stability. Generally speaking, a collective risk arises from the combined negative externalities of numerous actors (Vlek & Steg 2007). Negative externalities are related to the concept of 'public bads', where no one can escape the harm, as it is largely unknown to the persons affected, and the harm to one person does not reduce the harm to another. Externalities can hinder the implementation of collec-tive actions that could lead to optimal outcomes for a collective or a society as a whole. Externalities reflect the divergence between individual and collective interests. While it is in the collective interest, for instance, to provide a public good through collective action, it is in the interest of the individual to externalise harms that counteract the realisation of the goal of collective action, such as a public good. Furthermore, externalities manifesting at the macro level can require collective action, which must be coordinated to internalise them (Buchholz & Sandler 2021:535f.).

Often, externalities result from the choices of business models by private market actors to ‘outsource’ potential harms, rather than accidental technical failures. Factors of systemic risk from externalities can be differentiated according to the three idealised types of actor constellations and responsibilities: (1) through individual actors (providers or adversarial) or AI models with system-wide reach, (2) through interactions and coordination between many actors, such as collective action problems, or (3) through externalities that are incentivised or inappropriately constrained by an institutional arrangement. Externalities can take on a ‘systemic nature’, either as systemic externalities, through their interaction with other systemic risk phenomena (see below), via non-linear accumulation or due to feedback dynamics.

In the contexts that AI is implemented in, negative externalities are pervasive. Acemoglu considers the harms of AI to result from lacking regulation, assuming that competition does lead to their internalisation (Acemoglu 2024:663). Korinek and Balwit conceptualise externalities of AI as differences between societal goals and the actual alignment of values in the operation of AI systems. A socially aligned AI operation would then entail the internalisation of AI externalities (Korinek & Balwit 2024). The literature lists several externalities, which are caused by the way AI systems are provided, implemented and used, such as excessive automation, neglected or unforeseen impacts on income distribution as well as social and economics inequalities through deskilling or replacement (Acemoglu 2024; Korinek 2020; overview in Bengio et al. 2026:84-88), the outsourcing of low-status, low-wage annotation and content moderation labour to click-workers with bad working conditions and mostly in low-income countries (Hagendorff 2022), social costs by excessive worker monitoring (Acemoglu 2024), environmental externalities in form of emissions of greenhouse gases of data centres largely consuming fossil-fuel energy, water consumption for cooling as well as environmental damage caused by the extraction of raw materials for the hardware, which is primarily externalised to low-income countries (Hagendorff 2022; Domínguez Hernández et al. 2024; Schön et al. 2025). Furthermore, there are externalities of platforms’ business models to stimulate a long engagement of customers at the platform with AI-enabled feed-algorithms or recommendations, recommending and favouring polarising content, possibly adversely affecting the democratic functions (Korinek & Balwit 2024; see, in general, Summerfield et al. 2025; Gauthier et al. 2026). In the following, we discuss externalities caused by AI-based practices of informational emergence, proxy differentiation, as well as residual risks resulting from choices of trade-offs in AI development and deployment.

# 5 Phenomena that drive, influence and shape the systemic risks of AI

## 5.1 Informational emergence

AI-based data processing and decision-making give rise to a wide range of externalities. At first, there are externalities relating to privacy, which arise when the disclosure of personal data by one person reveals information about others, as in communications about third parties on social media platforms. A data controller can draw conclusions about the third party based on the data disclosed by others, which imposes a negative externality if the third party suffers harm, such as unfavourable treatments (MacCarthy 2010; Tucker 2019; Acemoglu 2024; de Brouwer 2020; Fainmesser et al. 2023; Choi et al. 2019).

AI can generate new data from existing data or contribute to its creation, resulting in 'informational emergence' (Drackert 2014). This phenomenon occurs as an emergent effect, since the associated risks cannot be explained by considering a single unit of personal data or a single instance of local data processing, but rather by the diverse interactions between the individual data processing activities and units (Esayas 2017). The concept of emergence is particularly relevant for data subjects (or persons affected, respectively), who, at the time they consent to the processing of their personal data, may have only a single instance of data processing in view, but may be largely unaware of the many other possible uses and their implications for them. The interaction processes that are decisive for informational emergence are data aggregation and merging, which then allows for the derivation of new information. They can also include reuse of data, changes to and extension of the purposes of data processing across actors and domains – especially when involving the repurposing or secondary use of personal data or a 'functional creep' (Koops 2021). The derived new information can be given in form of inferences, such as in predictions, profiles, scores, or rankings, and as part of decision-support and automated decision-making processes (Loi & Christen 2020; Solow-Niederman 2022; Lee et al. 2024). These can involve a single data controller or informational emergence can arise from the interactions of two or more organisational units such as private or public data controllers or data brokers.

These are not completely new phenomena,[7] but these problems are significantly exacerbated by AI, as it is particular capable of generating new information from existing datasets, such as its capabilities of pattern recognition (Solove 2025; Federal Constitutional Court 2023). The ability to generate new data from existing datasets is the central element of AI-based surveillance, be it by private companies (big tech companies, data brokers), governmental actors, or through the increasing interaction between these two groups (Viljoen 2021), leading to violations of fundamental rights and a loss of democratic control (Sharma & Adler 2026). AI is capable of drawing inferences about highly personal characteristics, biometric or sensitive data, even from trivial data, or of recognising or inferring emotional states (Rhoen & Feng 2018; Wright et al. 2026; McStay 2020) and political opinions (Kosinski et al. 2024). Furthermore, AI capabilities potentially enable de-anonymisation and re-identification (Rocher et al. 2019), such as the re-identification in anonymous datasets by identifying interaction patterns (Crețu et al. 2022). Such capabilities of AI can be used in mass personalisation, which can lead to mass persuasion (Matz et al. 2024; Susser et al. 2019) and mass deception (Park et al. 2024).

---

[7] They have been discussed in a similar vein, for instance in the context of data mining, profiling, big data analytics, or predictive analytics, sometimes under the terms of 'group privacy' or 'predictive privacy' (Hildebrandt 2008; Gandy Jr. 2010; Barocas & Nissenbaum 2014; Taylor et al. 2016; Mittelstadt 2017; Mühlhoff 2021). Similarly, these issues are addressed from a relational perspective on privacy, pointing out that personal data are interlinked insofar as decisions regarding the personal data of one person have implications for others and, in particular, that the disclosure of personal data by one person may reveal information about others (Humbert et al. 2019; Barocas & Levy 2020; Viljoen 2021; Solove 2022).

The AI-enabled identification and prediction of user preferences and behaviours through recommender systems can be used to control the selection and placement of messages, often with the aim of achieving incentives for the benefit of the platform, such as (long) engagement, site visits, and click-throughs (Ricci et al. 2021). Depending on optimisation criteria, this may lead platforms to overly emphasize emotional and polarising content (Brady et al. 2017; Rathje et al. 2021). It can also serve to exploit consumer surplus and to steer the composition of social media groups in a way that facilitates communication with like-minded users, leading to polarisation (Acemoglu 2024:667).

Another aggravating factor is that generative AI also poses privacy risks (e.g. Mireshghallah & Li 2025; Barberá 2025; Lee et al. 2024; Bengio et al. 2025:139-143), by processing much larger volumes of even unstructured and heterogenous data from different sources (including legacy datasets) at lower costs, making it a highly efficient tool for informational emergence. Informational emergence can already arise "within" generative AI systems (EDPB 2024), described as memorisation risks (e.g. Tonekaboni et al. 2025), enhanced capabilities to infer personal data from text processed by Large Language Models (LLMs) (Staab et al. 2024), or shadow profiling (e.g. Rohatgi & Park 2025), by combining individual pieces of personal data scraped from digital sources (e.g. the internet) to automatically generate comprehensive profiles.

It is also to be expected that these risks could increase dramatically with using non-public personal data, such as data from the interactions with and the use of generative AI products and services such as chatbots (according to the terms of service) (King et al. 2025) or from communication in social media, or with (anonymous) health data (e.g. Knolle et al. 2026).[8] In addition, the ongoing personalisation of AI applications in the form of personal AI companions and AI agents or based on the storage of communication traces or prompts in AI services leads to an extension of personal data, in particular sensitive data (Meurisch & Mühlhäuser 2021; FTC 2024).

With informational emergence, externalities (one causing actor or activity of data processing) or systemic externalities (more than one causing actor or activity of data processing) arise when persons affected do not have the ability to trace the full range of data processing, inferences and conclusions made, when their data is processed, transferred and (re-) used, inferences are drawn or proxies are generated, transferred and (re-) used or when personal data is stored in generative

[8] Among other things, the European regulation on the European Health Data Space (EHDS) allows the use of health data to train AI models (Article 53(1)(e) of Regulation 2025/327).

AI models. This preempts their ability to give meaningful consent to this infringement on their right to informational self-determination – that is, control over the processing of personal data relating to them and control over and shaping of the image or profile generated and (re)used about them. In such cases, they are deprived of the basis for knowingly participating in the benefits or receiving compensation for damages.

Such externalities extend considerably over time due to the persistence of digital data, the typical long-term storage of data, and the possibilities of generating new data and drawing inferences at points in time that are unimaginable by persons affected when giving their consent (Gellert 2020:217).

## 5.2 Proxy differentiation and bias

Two further types of potential (systemic) externalities are the bias-induced risks of discrimination by the use of algorithms and the encroachment on the right to receive explanations in order to give meaningful consent and to be able to influence decisions by the persons affected.

The algorithmic generation of classifications has macro-level implications for a society. Just like earlier technologies such as data mining and big data analytics, AI can be used to form algorithmically assembled groups that serve to classify individuals and make predictions and decisions at a group level rather than a level of individual assessments (Mittelstadt 2017), partly based on profiling and differentiation through the association and affinity of criteria ('affinity profiling') (Wachter 2020). This phenomenon is also discussed as 'proxy discrimination' or 'statistical discrimination' since differentiations are based on proxy variables built on statistical correlations (Schauer 2018).

Through the algorithmic generation of categories and patterns, data controllers draw conclusion by applying these patterns to individuals who may or may not be part of the original dataset (Solow-Niederman 2022), and potentially even without the use of personal data (Zuiderveen Borgesius 2016). The harms, particularly in the form of algorithmic discrimination and the perpetuation or amplification of bias and societal inequalities (Mittelstadt et al. 2016), go beyond the infringement of individual rights.

Externalities of discrimination arise when AI systems are deployed to differentiate between people, which imposes (residual) discrimination risks on affected individuals or groups. These affected persons are at risk of being treated differently without justification, while the provider or deployer reaps benefits, e.g. efficiency gains or cost reductions, from partial or fully automated decision-making (Barocas et al. 2022:Chp. 2, p. 15). These externalities can arise due to an insufficient debiasing of data and algorithms or because AI-models draw on dubious, sometimes pseudoscientific, relations between decision criteria and their goals (e.g. to grant loans) (Sloane et al. 2022; Andrews et al. 2024).

At the macro level, the characteristics of cumulative discrimination risks can be more serious than what can be analysed at the level of affected individuals. Causes are feedback dynamics from the reuse of biased and synthetic output data or data on behavioural changes induced by the outputs of AI systems (see Section 5.7). Furthermore, systemic externalities can arise from the transfer of biased proxies and inferences across applications, domains, and sectors.

Additionally, externalities of proxy discrimination are caused by reducing the judgement of individuals to quantified measures, using criteria and proxies to differentiate them instead of relying on individualized assessments and judgements (Britz 2008; Binns 2022). If the use of specific criteria and their weighting in AI-based decisions cannot be explained to the persons affected, they have no way of understanding the decision-making criteria, giving informed consent, making their position clear, and contesting or influencing the decision. Individuals are then treated merely as data objects, rather than being recognised for their individual personalities and free personal development. This can be understood as a disregard for their dignity (Teo 2023; Orwat 2024) and a threat to the rule of law.

### 5.3 Effects of technical design and trade-off decisions

Risks from the technical design of AI models and systems include the opacity of AI models, operations, and decision-making processes, misalignment between human intentions and AI operation in real-world environments, and the brittleness of AI systems when using out-of-distribution data (Gabriel et al. 2024; Amodei et al. 2016; Hendrycks et al. 2022). Bias is another source of risk and can lead to 'compounding injustice' at the macro level if the training data is inaccurate, selectively labelled, and perpetuates past injustices (Hellman 2023; Lippert-Rasmussen 2022). AI systems with a certain degree of complexity, speed, and technical autonomy can behave and fail in unexpected ways (Judge et al. 2024; Rhoen & Feng 2018; Maas 2018). Some authors point to inherent problems specific to generative AI and LLMs, such as implicit biases (Hofmann et al. 2024; Bai et al. 2024) and 'hallucinations' (Banerjee et al. 2025; Dumit & Roepstorff 2025). Even if AI models are initially highly accurate, their performance can degrade over time due to concept drifts or aging models (Vela et al. 2022).

There are also security risks and risks from interacting AI systems. AI could become a target of jailbreak and prompt injection attacks, data poisoning, and adversarial attacks, or be misused to generate malicious code and harmful content (Azzutti 2022; Arifovic et al. 2022; Amodei et al. 2016; Chan et al. 2023; Shavit et al.; Gabriel et al. 2024), with potential knock-on effects for the underlying IT infrastructure or cyber-physical systems. Furthermore, there are also systemic risks from interacting AI systems. Risks such as collusion between AI agents, harmful collective outcomes due to incompatible inherent rules, and new rules emerging from interactions between AI agents can accumulate. Simulations of multi-agent systems have demonstrated collective misalignment and the generation of collective bias that are absent in single models. Interacting AI systems can trigger unforeseen cascading failures and destabilising feedback loops for other systems, as seen in flash crashes or blackouts (Svetlova 2022; Chan et al. 2023; Gabriel et al. 2024; Hammond et al. 2025; Gazos et al. 2025; Min & Borch 2022; Darius et al. 2025). In the most extreme scenarios, AI swarms could be deployed by malicious actors to exploit vulnerabilities in democratic information ecosystems by creating a synthetic consensus and supporting the illusion that a majority agrees with strategically placed narratives (Schroeder et al. 2026). Ultimately, external threats and interacting AI systems can amplify existing risks posed by each individual AI module or collectively create new risks in sometimes unpredictable ways.

Technical design choices involve a multitude of trade-offs that can lead to residual risks, which can be understood as potential externalities imposed on the individuals affected. In predictive AI, predictions by AI systems are based on abstract goals set by human developers who try to grasp the issue accurately and find adequate measures, but rarely do so comprehensively and context-agnostically. Every technical design choice entails a set of expectations, while excluding others. Deciding on which data to use to train an AI ultimately faces the problem that social processes cannot be accurately predicted due to their non-deterministic nature. Therefore, the use of AI as a predictive tool in social settings is prone to distribution shifts (Wang et al. 2024), resulting in inaccuracies and errors. Social processes usually involve conflicts, compromises, and context-sensitivity; they are distinctly situated in specific spaces and change over time. Trade-offs also include those between the costs of risk prevention and mitigation measures and the risks of discrimination, between robustness and fairness (Li & Li 2024), between speed and accuracy (Dong et al. 2025), or between accuracy and energy savings (Mill et al. 2022; Schön et al. 2025).

Deciding on trade-offs also means choosing between efficiency and profit generation – which entails risk mitigation costs or the loss of revenue opportunities through refraining from risk-generating practices – and the (proportionality of) encroachments on fundamental rights or environmental protection (Orwat et al. 2024). However, the results of risk assessments, prevention, and mitigation measures by private actors, as well as their focus on efficiency and profit, can deviate significantly from solutions that are optimal for the public interest, particularly, when it comes to deciding on the acceptable levels of residual risks as potential externalities imposed on society by a handful of companies. Given finite resources, differing social expectations, and different understandings of what is risky or efficient, it can be expected that some residual risks will remain, even if the risk assessments and prevention measures implemented are state-of-the-art (Teo 2025). Residual risks remain as potential externalities that multiply with the number of uses of the AI model, system, or applications and can accumulate and manifest as systemic risks at the macro level.

### 5.4 Integration and coordination failures

In the following, the term 'integration' refers primarily to two things: integration along the value chain for the development and provision of AI systems, products and services; and the integration of AI systems and applications into other sectors or domains of society. In recent years, extensive interrelations have developed among a multitude of actors and technical components of AI with varying degrees of centralisation and risk distribution. Terms such as 'AI ecosystem', AI supply chain, or value chain illustrate that networks of actors coordinate their actions within interorganisational arrangements to achieve collective outcomes, either by jointly producing compatible products or by providing suitable services (Jacobides et al. 2021; Kak & West 2023; Brennan et al. 2025; Engler & Renda 2022; Hopkins et al. 2025a; Hopkins et al. 2025b; Fraser & Suzor 2025; André et al. 2025).

In these networks, systemic risks can emerge if multiple actors fail to coordinate their actions in a way that achieves the intended collective outcome, because they do not share knowledge in cases of information asymmetry, cannot agree on technical standards, or are unable to sufficiently align their activities. These problems of collective action produce externalities that may accumulate into systemic risks at the macro level. Previous research on information security, for instance, points to distributional issues, problems of collective action, and externality in decentralised markets, where economic incentives may lead individual organisations to misallocate resources, thereby not only endangering themselves, but also the security of a range of other actors (Anderson & Moore 2006; van Eeten & Bauer 2008; Bauer & van Eeten 2009; Moore 2010; Varian 2004; Kianpour et al. 2022).

There is also the phenomenon of repercussions resulting from collective risk prevention and mitigation efforts. These effects can be attributed to the complexity of multiple phenomena occurring simultaneously across different levels, driven by numerous interconnected actors with differing expectations. Furthermore, they are amplified by a multitude of infrastructure elements as well as hardware and software components of various provenances that are prone to generating unexpected feedback loops. Such repercussions can stem from well-intentioned patchwork interventions by upstream developers or downstream domain experts and users. Examples include fine-tuning with human feedback, models incorporating new information from external sources, or immediate optimisation techniques such as 'in-context learning' (e.g. Laufer et al. 2024, 2025; Williams et al. 2025).

The repercussions of mitigation efforts could also have organisational origins. The complex interactions between the countless technical and social elements mentioned above can lead to an organisational drift into failure modes – via "small, seemingly insignificant steps eventually contributing to extraordinary unforeseen events" (Dekker & Pruchnicki 2014:541). Even seemingly trivial matters, such as component degradation or a lack of maintenance, could lead to large-scale damage. The patchy integration and inconsistent operation of AI systems can lead to cross-sectoral and supra-organisational coordination problems. Without common goals, mediating actors, coordinated mitigation, and harmonised standards, multi-organisational networks are more likely to share vulnerabilities (Schulman 2023)

In centralised networks, such as the current 'AI ecosystems', these systemic risks could emerge if a central orchestrator fails to coordinate providers of complementary or derivative products and services, does not commit them to a common standard, or is unable to establish stable, shared interfaces through which other market members can interact and exchange information. Highly decentralised networks, on the other hand, might lack harmonised coordination or suffer from distributional failures due to free-riding by individual actors seeking to maximise their perceived short-term benefit (Jacobides et al. 2024), which then manifests as factors for systemic risks.

AI networks grow through the integration of foundation models into downstream products and services. As a result, they are increasingly in line with the description of complexity by Perrow: actors trying to mitigate risks within these networks receive information about interactions between sociotechnical elements only indirectly, via multiple monitoring and control devices with various control parameters that do not reflect the actual status of the elements. The elements inside these networks can also be seen as increasingly tightly coupled or interdependent, making them prone to accidents with unexpected and unintended feedback loops that can propagate rapidly and without the possibility to intervene (Perrow 1984; Sagan 1995). The increasing global availability of AI applications in different contexts, their inherent complexity, and their poor out-of-distribution performance exacerbate this problem, especially when more than one AI system is involved and interacts with others. The degree of tight couplings also corresponds to the degree of automation, brought about by organisations implementing AI into their processes to speed them up and increase performance or efficiency. Systemic risks could then materialise as failures cascading through the networks, especially when they are coupled with highly automated and time-critical operations (Maas 2018; Bianchi et al. 2023; Macrae 2022; Gazos et al. 2025). These could also take the form of 'data cascades', with data issues leading to compounding events that cause negative downstream effects (Sambasivan et al. 2021), algorithmic bias spreading through interconnected scoring systems (Chien et al. 2025), or implicit social scoring emerging in-between multi-agent systems (Darius et al. 2025).

### 5.5 Market concentration and network effects

The market formation around AI models, products, and services is currently concentrated and consolidating. Especially in the segment of LLMs, foundation models, or GPAI models, a handful of large tech companies hold a structurally dominant position. This concentration is driven by increasing resource requirements for training 'frontier AI' models (high fixed costs), (in many cases) privileged access to proprietary data, financial capabilities for costly acquisitions, and the scarcity of top talent. Moreover, strong learning effects and economies of scale arise when user interactions are tracked and utilized to improve AI system performance (Agrawal et al. 2018; Goldfarb & Trefler 2019; Varian 2019:412; Jacobides et al. 2021; Korinek & Vipra 2025; Bengio et al. 2025). The availability of large datasets to a few dominant providers is often rooted in strategies of massive data appropriation, which have been enabled, in particular, by inadequate data protection frameworks (Jacobides et al. 2021; Zuboff 2015).

Network effects (Katz & Shapiro 1994; Farrell & Saloner 1986; DiMaggio & Garip 2012), where the value of a network to a user increases with the number of other participants with whom they can communicate or exchange goods, can also consolidate the dominant position of platform providers using AI. These network effects could increase switching costs and lead to user lock-in. Further market concentration can result from 'walled garden' strategies, whereby platform providers facilitate the use of their own products or services and restrict access to competing ones. Due to the highly concentrated market, some platforms utilising AI systems can gain an enormous reach. For instance, Larsson et al. discuss 'LinkedIn' as an example of a far-reaching platform with implications for the labour market, such as exacerbating the risk of algorithmic discrimination (Larsson et al. 2024).

AI models can also have indirect network effects (Varian 2019:412) when integrated without modularity into downstream products and services developed by third-party providers. Even though a wider range of complementary products and services provides more utility for each contributor – hence an indirect network effect –, they also come with dependencies. This can not only cause the cascading failures mentioned above, but also path dependencies.

Market concentration usually implies less competition and innovation (e.g. products and services that might better respect fundamental rights or are more environmentally friendly), narrows the scope for experimenting with alternative solutions for risk prevention and mitigation, and limits the choice of models for downstream AI providers or users. The latter can lead to greater dependencies and less favourable contractual terms, or even to unilateral dictates. Structurally dominant actors can shift the responsibility for prevention and mitigation to downstream providers, deployers, or users (Hopkins et al. 2025b), leading to an externalisation of risks. Furthermore, with regard to systemic risks in finance, Linarelli points to the regulatory threat of too-big-to-fail situations arising from concentration, in which companies may acquire unjustified subsidies (Linarelli 2017). Given the current market structure, this threat may also arise in AI contexts.

### 5.6 Algorithmic monocultures, outcome homogenisation, and narrowed diversity of output

In algorithmic monocultures, the same model is applied to many consequential decisions (Kleinberg & Raghavan 2021; Creel & Hellman 2022; Bommasani et al. 2022). Outcome homogenisation refers to scenarios in which an individual or group receives negative results across different AI systems because these models rely on similar training data or the same foundation models (Bommasani et al. 2022; Creel & Hellman 2022).[9]

Additionally, due to their performance advantages, GPAI or foundation models are likely to be adopted across a vast range of consequential tasks (Creel et al. 2021:152). If a few homogeneous GPAI or foundation models are integrated downstream into other AI systems the inherent failures of the GPAI model can propagate across a wide range of these systems. This outcome homogenisation can lead to the systemic exclusion of certain individuals or groups (Bommasani et al. 2022; Creel & Hellman 2022). In the job market, for example, such systemic exclusion can manifest as the complete rejection of a jobseeker by all potential employers, rather than just a handful of rejections by individual decision-makers, leading to a non-linear systemic failure (Toups et al. 2024:2). In another instance, a study regarding the use of hiring algorithms to screen and rank job applicants provides evidence for systemic rejections based on race and at scale (Bommasani et al. 2026).

Related to these issues are phenomena described as reduction of diversity in output (Matz et al. 2025; Ghafouri 2025), 'variance reduction' (Xie & Xie 2025), 'generative monoculture' (Wu et al. 2024), or the homogenisation of language, perspectives, opinions, and 'reasoning' in LLMs (Sourati et al. 2026). In these scenarios, generative AI models tend to produce an 'average' outcome characterized by reduced diversity and delivering output more 'around the mean'. If a small number of actors leverage these low-diversity LLMs to influence populations in their interests, it could pose a systemic risk to democracy (Sourati et al. 2026).

---

[9] Other authors point to the multiplicity of models typically available for a given task and to the multiplicity and the inherent arbitrariness of algorithmic predictions (Gur-Arieh & Lee 2025; Gorecki & Hardt 2025).

Algorithmic monoculture, homogenisation, and reduced model diversity can also mean that the errors of the few remaining models can spread widely throughout ecosystems, supply chains, or value chains, while downstream organisations may lack the means to mitigate them (Hopkins et al. 2025b). Furthermore, the reduced diversity of biased outcomes can promote stereotypes and perpetuate or amplify algorithmic discrimination when models are deployed in consequential decisions. Overcoming algorithmic monoculture and outcome homogenisation constitutes a collective action problem (Bommasani et al. 2022:10). Individual deployers may lack the incentive to replace a model due to its performance advantages or its tight integration into their AI systems, yet systemic discrimination risks inevitably emerge from their parallel behaviour.

### 5.7 Feedback dynamics and feedback loops

While the concept of feedback has a long and varied history within the social sciences, definitions generally converge on the notion of a closed causal loop. These self-reinforcing interactions between system elements or characteristics either stimulate divergence from an initial state (positive or self-reinforcing feedback) or stabilise it (negative self-dampening or stabilising feedback) (e.g. Arnscheidt et al. 2025; Lawrence et al. 2024; Richardson 1991). In the extreme, positive feedback can push a system across a tipping point, causing a transition towards an alternative stable state, equilibrium or 'basin of attraction' respectively, whereas negative feedback can result in path dependencies or stasis. These dynamics are generally considered a defining characteristic of complex systems and at least one of the reasons why their behaviour is difficult to predict and govern (Ferrara 2024; Kolt et al. 2025). Although crucial to the non-linear behaviours of complex systems, feedback dynamics are not necessarily problematic and can even be desirable, for instance, when facilitating tipping dynamics to accelerate the transformation towards carbon-neutral societies (Geels & Ayoub 2023; Otto et al. 2020; Lawrence et al. 2024).

From the perspective of systemic risks, feedback dynamics are relevant in at least three ways. First, undesirable positive feedback dynamics can drive a system towards a suboptimal fulfilment of societal goals, thereby threatening or destabilising their emergence. Second, undesirable negative feedback dynamics can stabilise an inferior system state, preventing or hindering the realisation of these goals. Third, desirable feedback dynamics that support or stabilise the emergence of societal goals may be actively threatened. In each case, externalities may be amplified (Fleurbaey et al. 2021; Nokhiz et al. 2025).

In the context of AI, various feedback loops have been observed. So far, most are mediated by behavioural responses to AI system outputs and data. Model decisions influence behaviour, which subsequently affects outcomes that then re-enter the decision context as training or observational data, making similar decisions more likely (Glickman & Sharot 2024; Hardt & Mendler-Dünner 2025). For example, predictive policing systems may identify a certain area as crime-prone. Whether the initial judgement is biased or not, this can lead to increased policing, yielding a disproportionate recording of new crimes and reinforcing the initial judgement (Ensign et al. 2018; Lum & Isaac 2016; Richardson et al. 2019). Comparable effects have been documented in refugee status determination (Kinchin 2024) as well as in hiring processes (Fabris et al. 2025; Kim 2016). Similarly, recommender systems can induce self-reinforcing behaviours or shifts in belief by generating content, tracking user responses, and tailoring subsequent suggestions accordingly (Pedreschi et al. 2025), often prioritising engagement over user interest (Ekstrand & Willemsen 2016; Li et al. 2024; Ricci et al. 2021), resulting in harms to political processes of democracy.

Some feedback loops mainly occur based on interactions and model training, sometimes also including a behavioural component. For example, without intervention, new AI models increasingly fail to represent reality because synthetic data generated by the own or other models is scraped and ends up in the training data. This phenomenon of 'model collapse' occurs through a recursive compounding of statistical and data processing problems. As current model generations build on previous generations, and a substantial amount of generated data, the real-world datapoint distributions are at risk of being distorted further (Shumailov et al. 2024). As a related phenomenon, synthetic data is also prone to reinforce bias in a potential feedback loop. As training datasets contain an increasing proportion of synthetic content, errors and biases compound over time, resulting in a 'fairness feedback loops' (Wyllie et al. 2024; Taori & Hashimoto 2023).

Notably, these identified feedback dynamics tend to involve relatively straightforward causal loops encompassing a modest number of system elements. However, particularly in large-scale complex systems, various interrelated feedback dynamics may co-exist, involving many system elements and cross-boundary influences (Lawrence et al. 2024). While these dynamics are comparatively harder to identify due to more dispersed and contingent causal chains, it is likely that unknown problematic feedback dynamics are as relevant to systemic risk as the known instances described above. It stands to reason that many feedback dynamics relevant to the emergence of societal goods and affected by AI have yet to be uncovered. For example, other phenomena in this study, such as interacting AI systems or network effects, may be subject of certain types of feedback dynamics.

## 5.8 Systemic environmental impacts and rebound effects

Literature on the environmental impacts of IT and AI frequently distinguishes between first-order effects – directly related to production, use, application, and disposal – and second-order or systemic effects. The latter occur due to potentially unanticipated and unintended interaction effects between elements of a sociotechnical system, which in turn affect the environment and may threaten the societal goal of sustainability (Kaack et al. 2022; Rivera et al. 2014). 'Rebound effects' represent a prominent example of problematic systemic environmental impacts[10] (Luccioni et al. 2025; Kaack et al. 2022; Rivera et al. 2014) and have already been documented in the context of AI (Ertel & Bonenberger 2025; Wright et al. 2025).

Specifically, the concept of the rebound effect, also known as 'Jevons Paradox', describes a situation in which an improvement in the relative resource efficiency of a good, service, or function fails to achieve the expected absolute reduction in resource use, or even backfires, leading to higher absolute consumption. This outcome is driven by shifts in consumer behaviour in response to efficiency gains, mediated by economic, social, and psychological mechanisms. Similar to network effects, rebound effects are evolutionary processes that play out through the market-based interactions of certain system elements, including efficiency innovations as well as the resources embedded within goods, services, or functions (e.g. Vivanco et al. 2022; Vivanco et al. 2016; Giampietro & Mayumi 2018; Santarius & Soland 2018).[11] They can occur across multiple dimensions, such as different resource types (e.g. energy and water), or through shifts in how consumers allocate time, physical space, and their available income (e.g. Vivanco et al. 2022; Luccioni et al. 2025).

[10] Other examples of systemic environmental impacts may include the potential to enable pro-sustainable innovations, induction and scale effects, changing practices (Rivera et al. 2014; Rolnick et al. 2022), or AI-related path dependencies (Kaack et al. 2022).

### 5.9 Information asymmetries

Information asymmetries, which exist when different actors possess varying scopes and qualities of information, are pervasive in the multi-layered relations between the AI system, the original model providers, downstream providers, deployers, users, and persons affected. They originate from the opacity of AI models and algorithms as well as from attempts by businesses to protect their secrets or intellectual property (Burrell 2016; Kroll 2018; Kim 2020). These asymmetries *de facto* relate to a broad range of activities surrounding personal data (collection, transfer, selling, and reuse). They encompass both the activities of diverse actors (such as data controllers, brokers, and intermediaries) and the consequences for affected individuals, including the drawing of inferences, the generation of profiles, as well as their transfer and (re-)use, i.e. the activities resulting in informational emergence. The rules applied for categorising and differentiating persons, the actual implementation of data protection principles (such as purpose limitation), and the true effectiveness of consent also play a critical role. Information asymmetries can increase alongside the decision-making autonomy of AI-based systems, such as AI agents. Additionally, the personalisation and adaptive behaviour of AI systems contribute to these asymmetries, as they require vast amounts of personal data, making it in turn harder for affected persons to evaluate personalised offerings relative to other individuals or across different points in time.

In general, information asymmetries can exacerbate disparities in contractual relationships regarding the provision and use of AI. In many cases, contractual relations become the vehicle through which fundamental rights or sustainable development are actually realised. These asymmetries between contracting parties typically lead to 'lemon market' or 'moral hazard' problems. The alignment of AI systems – such as compliance with fundamental rights or environmental friendliness – can be understood as distinct qualities of AI-based products and services. In 'lemon market' situations (Akerlof 1970), the demand for offerings with these qualities, and subsequently the investment into their production, can fall well below the socially optimal level. This happens if these qualities cannot be communicated on markets in a trustworthy manner, for instance, due to a lack of generally understandable, acknowledged, and trustworthy indicators or labels. For example, without reliable information about the energy consumption of each prompt, sustainability cannot serve as a criterion in buying or use decisions.

[11] Producers or providers of alternative or tangential goods, services, and functions may also be involved in generating such effects, specifically due to substitution or induction effects (Luccioni et al. 2025). However, applying the concept of rebound effects so broadly as to include these kinds of dynamics can pose conceptual challenges (see Vivanco et al. 2016). Irrespective of this, they must be considered when evaluating systemic environmental impacts.

In 'moral hazard' situations (Arrow 1985), actors with an information advantage within contractual relations tend to engage in opportunistic behaviour and are inclined to ignore all negative consequences of their actions, i.e. to externalise risks to other parties. In AI supply chains, for instance, information asymmetries decrease the explainability and traceability, thereby limiting the options to attribute responsibility for causing, preventing, and mitigating risks (Hopkins et al. 2025b).

In addition, information asymmetries make it difficult or impossible for those affected to recognise and address the violation of their rights and enforce them. This includes, for example, restrictions on the right to privacy and data protection that take effect with a time lag after data collection. It also encompasses discrimination through the withholding of offers or through personalised offerings, resulting in individualised treatments that are difficult to compare with how others are treated. If affected individuals lack vital information, they remain at a structural disadvantage when taking personal action against these restrictions. Furthermore, through the accumulation of personal data and the creation of personal profiles information asymmetries can leave individuals vulnerable to manipulation via targeted information designed to influence their voting behaviour, shape their political preferences, or encourage them to purchase goods. They also hamper the ability of persons affected, researchers and the broader public to investigate, contest and verify the causal relations between differentiation goal (e.g. to differentiate the creditworthiness) and applied criteria (or vectors of criteria). This can encourage the use of dubious correlations and pseudoscientific criteria when individuals are categorised, even if the assessment is accurate according to technical benchmarks.

Information asymmetries can also cause chilling effects. They describe how persons affected by surveillance, profiling and practices of informational emergence are deterred from enjoying fundamental rights, like the right to free speech (Federal Constitutional Court 1983) or from consuming products or using services (Büchi et al. 2020). Chilling effects can arise, when persons (potentially) affected are uncertain about the actual scope and severity of surveillance, the generation and processing of personal data and the inferences and decisions made about them. At the societal level they can emerge as systemic externalities resulting in distrust or disfunction of democratic processes.

Multiple information asymmetries and a lack of traceability in cause-and-effect relations restrict the capacity to resolve collective action problems and manage complex externalities (see Section 4.2). In scenarios requiring collective action for the provision of collective goods, these asymmetries can obscure the interests and rationalities of providers, deployers, and affected parties, which usually diverge greatly. Consequently, they hinder or prevent self-organised collective actions, as the contributions of individual participants are unclear or not apparent to others. This makes it more difficult to establish self-governed mechanisms of reciprocity or sanctions in the event of missing or poor contributions. In situations with aggregate and complex externalities, information asymmetries impede the ability to estimate the size and severity of externalities, thereby making self-organised bargaining solutions between risk-causing and risk-bearing actors more difficult. The difficulties caused by these asymmetries can rule out decentralised negotiation solutions by market actors alone.

### 5.10 Normative ambiguities

As emphasised by Renn et al. (2022), normative ambiguities are one of the properties of systemic risks, becoming apparent in differing and sometimes conflicting interpretations of risk-related facts (see also Johansen & Rausand 2015). Especially for AI, normative ambiguities in interpreting and operationalising fundamental rights, determining levels of acceptable (residual) risks, and deciding on trade-offs, value conflicts, and the proportionality of fundamental rights infringements hamper the identification, assessment, prevention, and mitigation of risks (Orwat et al. 2024).[12] The resulting uncertainties in the regulation of AI can be a source of systemic risk itself, since they diminish the effectiveness of regulation and the protection of fundamental rights. Risk assessments imposed on businesses by regulation can lead to ambiguous results, and thresholds for acceptable risks may be absent or appear arbitrary. The consequent inappropriate normative decision-making in development and use of AI systems and applications (e.g. about trade-offs and residual risks) can systematically lead to externalities.

Systemic risks can arise if it remains unclear which normative value system is applied in risk assessments and risk management – for example, if specific forms of AI safety or AI ethics originating from AI providers prevail over fundamental rights, if value systems are not integrated coherently, or if the applied normative value system does not align with or achieve the provision of societal goals. Implementing unsuitable value systems, interpretations, and operationalisations in benchmarks, key performance indicators, metrics, or guidelines creates flawed system-wide incentives for the development and application of AI systems. For instance, Watkins and Chen demonstrate that fairness metrics built on the so-called 'four-fifths rule' often apply this rule inappropriately. Without embedding it into the other essential rules of anti-discrimination law, such isolated applications lead to epistemic trespassing (Watkins & Chen 2022).

### 5.11 Deficits in the institutional and regulatory framework

Deficient institutional arrangements represent a major source of systemic risks. These include legal or regulatory obligations that provide inappropriate incentives when deciding on trade-offs and acceptable levels of residual risk imposed on society. Such deficits are further compounded by institutional incentives to externalise risks as well as gaps, failures, or legal but inappropriate exceptions within the regulatory framework that de facto enable risk-causing behaviour. However, these regulatory dimensions can only be briefly touched upon here.[13]

[12] For a recent study on these challenges in the context of the European AI Act, see Yeung (2026), who argues that unless these challenges are tackled, the risk management required by the AI Act will become a 'theatres of compliance'.

Several elements of the current European regulatory framework are discussed as such in view of AI risks. In general, the individual rights approach – meaning rights that guarantee affected individuals the legal option to correct or compensate for harm – is unable to cope with system-wide failures and systemic risks. Affected persons may be increasingly overwhelmed by the enforcement of their individual rights, largely due to the plethora of cases involving data processing and algorithmic differentiation that are difficult to manage. Moreover, systemic harms can simply be unrecognisable to those affected. For instance, inferences about groups can be drawn with AI applications even when no personally identifiable data about specific individuals is processed (see Section 5.1). Privacy laws primarily require that information be provided to affected persons about the gathering of personal data, but not about the generation of new personal data and the drawing of inferences based on the gathered data (Solove 2025:37).

Other elements of governance frameworks have lost their effectiveness in view of AI-induced harms, not at least with the current deregulation efforts within the EU regulatory framework. In data protection law, this includes weaknesses in the informed consent approach or the principles of data minimisation and purpose limitation – especially with regard to secondary data use for AI training (Mühlhoff & Ruschemeier 2025) – a too narrow interpretation of what personal data is and, thus, what is covered by data protection obligations, and an overly broad interpretation and use of the 'legitimate interest' provision for processing personal data without the consent of affected persons. Furthermore, it is questionable whether the governance elements of 'regulated self-regulation' in the form of risk assessments by AI providers and data controllers in the GDPR and the AI Act lead to outcomes in the public interest. The latter demands a maximisation of the realisation of fundamental rights (Almada & Petit 2025), rather than the regulatory balancing of fundamental rights protection against the private costs of risk prevention and mitigation or the loss of private profits. It is therefore unclear if the cumulative impact of self-regulation also results in an erosion of fundamental rights at the macro level.

Furthermore, discrepancies frequently arise in the application of the proportionality principle, if existing balancing practices may not be adapted to the new technical capabilities of AI. For example, this occurs when the data of a vast number of subjects is processed with AI-based surveillance only to identify a few individuals (EDPB/EDPS 2021:12). Within the regulatory frameworks of the AI Act and the GDPR, these balancing decisions are mainly made by AI providers within their risk assessments and risk management (Orwat et al. 2024).

Additionally, a creeping erosion of fundamental rights, notably the right to human dignity, occurs with the transition to automated decision-making based on complex AI systems, such as LLMs or predictive systems based on deep neural networks. The erosion occurs when the logic involved in decisions – including the criteria or their weightings – can no longer be explained.

---

[13] See for further analyses on the GDPR (Li et al. 2025), on anti-discrimination law (Wachter et al. 2021; Orwat 2020), or on the AI Act and liability regulation (Wachter 2024).

The provision of explanations of the involved logic serves the right to human dignity (Teo 2024; Orwat 2024).

Other weaknesses within current regulatory frameworks include inconsistencies in legal provisions and a fragmented assignment of responsibilities to different authorities, causing pervasive legal uncertainties (e.g. Wachter et al. 2021:8; Greenstein & Zamboni 2025:34-35). This normative and legal uncertainty disadvantages small and medium-sized enterprises and start-ups with fewer resources to interpret complicated legal provisions, efficiently conduct risk assessments, prevention, and mitigation, or litigate for the clarification of regulatory ambiguities, ultimately stifling innovation and competition. Another normative ambiguity is the lack of societal agreement regarding acceptable thresholds for systemic risks. This deficit often results from disagreements over future states of sociotechnical or ecological systems and the resulting conflicts of interests.

# 6 Discussion

## 6.1 Interacting phenomena of systemic risk

The distinction between the aforementioned phenomena of systemic risks is analytical. In effect, the phenomena can overlap, interact and amplify each other, and in their interaction provide possible patterns of causing systemic risks. This has three implications for assessing, preventing and mitigating systemic risks.

First, many of the phenomena are interdependent and influence one another, as well as the manifestation of systemic risks and their severity. Therefore, the causes and impacts of systemic risks can hardly be explained by considering individual phenomena in isolation, but only through the complex interactions of multiple causes and effects, including the fact that some effects may also be causes of other risks (Simpson et al. 2021; Gambhir et al. 2025). For instance, the interaction of market concentration, integration, network effects, information asymmetries, and the resulting structural dominance of some AI system providers or deployers has an amplifying effect on risks to fundamental rights, as affected persons may find it considerably more difficult to realise or enforce their own rights. This is not only due to their greater dependence on the product or service, but also due to the reduced range of available alternatives to avoid products or services that are perceived as possibly harmful. Furthermore, information asymmetries and personalised offers may make it less clear if and in what way persons affected are actually at risk and which harm should be challenged. The overall structural dominance of some AI system providers or deployers in the aforementioned scenario may come close to the level of state actors. In this scenario, the same standards of fundamental rights should also apply to structurally dominant private actors.

Second, the actual occurrence and dimension of systemic risks usually depend on the specific interaction of configurations, processes and interactions between the aforementioned phenomena in particular contexts – that is, within specific social or global systems. Whilst many such interrelations between systemic risk phenomena are conceivable and plausible in conceptual or theoretical considerations, their actual interaction, appearance and dimensions of systemic risks can depend on particular actual interactions and behaviour of the system elements in those contexts. For example, if rebound effects are part of a feedback loop of increasing consumption is contingent upon whether technologically achieved efficiency gains by scale and learning effects that may further stimulate demand by price reductions coincide with a latent demand potential. Additionally, increased revenues may or may not unlock further efficiency gains through R&D or process optimisation. A further example is the influence how measures of risk prevention and mitigation are actually implemented in practice. For instance, the actual appearance of feedback dynamics that can reinforce discrimination or inequalities depend, among other things, on the debiasing measures by providers and deployers that are actually put in place; or systemic privacy risks depend on the practices of informational emergence that are actually applied. It also becomes clear that the institutional and regulatory framework, which is actually effective in certain contexts, must also be taken into account, as it sets the incentives and constraints for the measures actually implemented and the extent to which risks are mitigated.

Third, the constellations of actors, their responsibilities (see Section 2) and their interactions within the sociotechnical systems from which systemic risks emerge are subject to constant change, and with them the potentially contributing phenomena described above. For instance, a further market concentration and the integration of value chains around dominant actors would reduce the number of actors involved, meaning that single dominant actors significantly contribute to the emergence of systemic risks. Additionally, with increasing government regulation of collective action problems – such as, for example, through information requirements in the relationship between GPAI model providers and downstream providers – regulation has to be increasingly considered as a major contributor to systemic risks.

### 6.2 Risk accumulation, non-linearities and acceptability

Emergent phenomena can result from the accumulation of individual risks, for instance in form of a general decline in privacy protection, a general increase in discrimination, or a generally inadequate level of security. The accumulation of risks can lead to non-linear effects if the severity of the overall risk level disproportionately increases as risks accumulate. This can happen when the state of a societal or global system changes beyond certain thresholds or limits, for example when a fundamental right in a society must be seen as no longer protected.

Non-linear effects can also be found on the side of those affected, in terms of exposure, their vulnerability or capacities to respond (Simpson et al. 2021). For systemic discrimination, for instance, exposure increases in a non-linear manner, if AI systems for the differentiation of persons are deployed in key application areas at an exponential rate. This is particularly the case if a small number of biased GPAI models are operating in a large number of IT systems without any downstream adjustments, and in this way generate a significant amplification effect. New vulnerabilities are generated if new AI-generated vectors or algorithms to categorise and differentiate persons – not contestable and verifiable by persons affected or by society – are not covered by existing anti-discrimination laws. The scale of damages from a certain discrimination harm or amount of exposure may tip beyond a certain threshold, overwhelming the capacity of a person affected or a social system to respond and mitigate harm. In the case of increasing market concentration or algorithmic monoculture, persons affected by rejections of applications by the same or an equivalent system deployed throughout the labour market, for instance, are more exposed and have fewer means to respond by evading to alternatives, so that a non-linear progression of damage can be assumed (Toups et al. 2024). In general, the erosion or lack of means of response can be understood as a stressor to the overall system, which then leads to a trigger yielding non-linear effects (Lawrence et al. 2024).

These phenomena present a challenge of determining acceptable levels of risk accumulation for a society or at the global level and determining at what point the accumulation leaves these boundaries becoming systemic. The scale of systemic risks at the macro level cannot be determined by assessing individual risks for each provider at the micro level and then aggregating them at the macro level. Instead, risks interact as aggregate, compound, mutually exacerbating or mitigating risks, so that the overall systemic risks at the macro level are likely to deviate from the sum of all individual risks. The level of overall systemic risks deemed acceptable to a society can only be determined from a societal perspective by assessing the degree to which the societal goals are realised. Since the assessment, balancing, and determination of acceptable risks or restrictions to fundamental rights by the AI providers themselves is very likely to deviate from the level required in the public interest, this balancing necessitates public institutions. Inappropriate trade-off decisions by private actors can multiply the risks posed by the many large-scale AI applications, a situation further exacerbated by the concentration of AI markets, algorithmic monocultures, and information asymmetries. Not or inadequately determining the levels of acceptable risk poses a systemic risk by itself. Formal efforts to reduce risks remain ineffective as long as the societal level of acceptability is not agreed upon.

### 6.3 Path dependencies

Path dependencies can be understood as manifestations of systemic risk processes over time. They can become problematic when a relatively stable sociotechnical development path proves to be inferior to alternative paths. In this way, externalities are imposed on future generations.

Path dependencies occur as emergent phenomena at the macro level (Arthur et al. 1987; Page 2015:28f.) and can be caused by different forms of feedback or self-reinforcing dynamics (Dobusch & Schüßler 2013; Pierson 2000). Among the factors are increasing returns to adoption, network effects, or learning effects from the use and improvement of a technology (David 1985; Arthur 1994; Liebowitz & Margolis 1990; Dobusch & Schüßler 2013). When technology diffuses, learning processes take place, leading to technological improvements, which in turn increase the likelihood of subsequent users adopting the technology. Learning effects can be complemented by specific investments in certain techniques or technical solutions, resulting in 'sunk costs' for equipment, tools and the associated skills required to use them. This may be the case, for instance, when GPAI models are integrated into downstream AI systems in a non-modular way. Path dependencies can result from complementary investments together with developments towards algorithmic monocultures or reduced diversity in AI research and development (Mateos-Garcia & Klinger 2023). Additionally, when AI is used in platforms' products and services, network effects may arise which make it seem more advantageous for users not to leave the platform. Network effects occur, for example, when platforms are able to collect more data, which they can then use to improve their products and services – such as more accurate AI-based recommendations (Mateos-Garcia 2018; Mateos-Garcia & Klinger 2023). In addition, path dependencies could become relevant if the AI industry increasingly takes on the characteristics of an infrastructure upon which other industries or sectoral systems rely, in the sense that the latter integrate AI applications into their practices and processes, leading to 'lock-in' situations (Robbins & van Wynsberghe 2022).

Among the various explanations for institutional path dependencies, the one of North emphasizes, among other things, the existence of costs of institutional change and the fact that interest groups may have incentives to defend an institutional status quo they profit from, i.e. if they have developed skills to operate within specific rules (North 1990). Similar aspects are conceivable in the AI context, such as stabilising effects for certain conceptual approaches in AI regulation (e.g. the narrow, but dominant concept of 'high-impact capabilities' as a narrow, but dominant understanding as causing systemic risks) or the use of certain metrics for risk assessments, risk mitigation and governance, even if they turn out to be inappropriate to achieve and safeguard societal goals. Providers and deployers may also try to leverage their potential structural dominance and political capital in order to facilitate the emergence of a path dependent system trajectory that benefits them.

# 7 Challenges and open questions

Since research into systemic risks of AI is still in its early stage, there is a need for further research, including adequate methods to identify, assess or measure systemic risks as well as options for preventing or mitigating them. Further conceptualisations are needed on micro-macro processes of emergence of systemic impacts on human rights or fundamental rights including non-linearities in the severity of violations of fundamental rights, that can result from amplification and feedback dynamics, but also from the accumulation of risks, cumulative impacts and compound risks of different interacting risk factors.

Research is also needed on the impacts of AI (e.g. biased or stereotypical outcomes) within complex social interactions in form of social influences, e.g. how shared expectations, opinions and beliefs developed among peers, family members, relatives or other members of social groups, in effect shaping the choice of actions and interactions, which can alter, amplify or perpetuate societal inequalities, violations of fundamental rights or derogations of democracy in systemic processes, for instance, with feedback dynamics or network effects.

Although there are initial insights into the nature of tipping points in social processes (e.g. Juhola et al. 2022; Jagers et al. 2020; Abrams et al. 2026), the tipping points at which processes are triggered that, taken together, lead to unacceptable degradations of protection levels of fundamental rights or even violate the essence of fundamental rights have not yet been sufficiently researched. However, knowledge regarding tipping points, feedback dynamics, and non-linear processes is necessary to define the overall acceptability of AI applications and their systemic risks as well as to justify and design governance interventions. For instance, insights from applying complexity science to social inequalities and discrimination suggest that interventions should be adequate to leave the 'basin of attraction' of systemic forces (Durlauf et al. 2025). There has also not yet been sufficient research in the field of AI applications to determine when the above-mentioned phenomena, in particular, relating to feedback and amplification patterns and path dependencies in sociotechnical developments become stresses to the stability of social or global systems, in the sense that they increase instability to the point where a trigger event can lead to crises (Lawrence et al. 2024), but also in the sense that inferior development paths can be abandoned or new paths can be created. This includes the differentiation between unwanted instability that may lead to crises and socially desirable 'instability' of inferior system states that can give way to necessary transformations, e.g. towards sustainable development or leaving paths of growing social and economic inequalities.

The concentrated AI market, the tightly coupled integration, network effects, the opacity of foundational models and information asymmetries on practices of informational emergence are prone to the phenomenon of structural dominance of a few actors. This raises research questions about capabilities and vulnerabilities on the side of affected persons and groups, in particular the distribution of (systemic) risks and benefits (e.g. Linarelli 2017; Hansson 2013), the options for persons affected to choose alternatives and means to understand, respond, contest and correct AI-based treatments (e.g. Svetlova 2022:716), and if this reinforces injustices (e.g. Behrendt & Loh 2022).

Outcome homogenization and narrowed diversity of output may lead to a standardisation of human characteristics and behaviour in AI-based assessments and decision-making to distinguish between persons. The systemic implications for the overall scope and qualities of opportunities for free personal development, self-determination, equality, or collective political development and participation can be conceived, but have yet to be adequately researched, particularly in systemic contexts. The application of such systems by actors with structural dominance, such as state actors or private actors with a dominant market position, and the increasing reliance on a few AI systems and models are particularly relevant as potentially amplifying phenomena.

## 8 Conclusions

According to the focus of this paper on emergence as a key feature of complex systems, the realisation and protection of various societal goals, such as fundamental rights, democracy and sustainability, are conceptualised as emergent properties at the macro level threatened by various forms of interactions among sociotechnical elements (e.g. humans and AI systems) at the micro level. Systemic risks of AI are conceptualised as the interplay of risk phenomena that can threaten or even entirely impede the realisation and protection of these emergent properties at a societal or global level. Among these social processes, collective action problems and the generation of externalities in complex situations are seen as the main contributions to the emergence of systemic AI risks. The self-organised solutions of complex externalities and collective action problems can be hampered by the aforementioned complexity features and phenomena such as normative ambiguity and heterogeneity of interests and capabilities, several forms of feedback dynamics, information asymmetries, or structural dominance by some actors. An understanding of these phenomena and factors can therefore also serve to justify and shape the governance of systemic risks.

In contexts of AI applications, collective action problems can arise in different forms. In increasingly integrated AI value chains, coordination failures in joint risk mitigation – for example, between GPAI model providers and downstream providers of AI systems – constitute a basic problem of collective action, which can lead to failures or risks to fundamental rights propagating to downstream applications. Furthermore, with market concentration or algorithmic monocultures existing and with the integration of AI systems into a multitude of economic, social, governmental practices and a multitude of domains or sectors, the collective action problems regarding a small number of models and AI systems give rise to a significant amplification effect. In this respect, collective action problems can lead to systemic externalities resulting from failed coordinated measures to mitigate AI risks to safety, privacy, information integrity, anti-discrimination, and other values. Systemic externalities can also result from the multitudes of data processing practices leading to informational emergence, i.e. the use of merged data to generate new data and inferences from existent datasets as well as spillovers of inferences for the differentiation of persons (e.g. profiles and proxies), across applications, domains, or sectors, far beyond the control of affected persons and without their knowledge. This phenomenon is exacerbated by information asymmetries and in case of structural dominance. Usually, such social processes are regulated by the institutional and governance framework. However, shortcomings of this framework and of related risk mitigation measures have also contributed to the emergence of the systemic risk phenomena themselves.

Framed from the perspective of complexity, systemic risks as emergent effects in complex sociotechnical systems cannot be causally reduced to these aforementioned phenomena in isolation. Nonetheless, they can serve as analytical heuristics and plausible points of intervention to mitigate risks and create opportunities for distributed actors in critical positions to identify when risks begin to emerge. Based on this understanding of systemic risks, it can be concluded that conventional approaches and methods of risk assessment are insufficient for assessing systemic risks (already Lucas et al. 2018:293). In the context of AI, the conventional approaches would focus primarily on assessing and optimisation of individual AI systems, suggesting that responsibility lies with individual AI providers. In contrast, the identification, analysis and assessment of systemic risks must take into account the diverse forms of interaction and connections between the system elements, the multifaceted social and sociotechnical processes through which societal goals emerge as well as the phenomena that give rise to systemic risks that hinder the achievement of societal goals. The comprehensive identification and assessment of systemic risks is more adequately achieved by using several complementary approaches, i.e. by distributed but responsible actors – such as AI providers, deployers, or users with exclusive access to relevant knowledge about sources, types and dimensions of risks – in conjunction with institutions who focus on the interacting and aggregating processes that ultimately give rise to systemic risks. However, issues of responsibility for assessing and mitigating systemic risks needs clarification. Certain actors like AI providers and governing institutions are, by virtue of their centrality to the sociotechnical system and resources at their disposal, in a position that affords them considerable capabilities with respect to systemic risk assessment and mitigation. Nonetheless, other actors within the system can, by virtue of their own positionality, offer complementary contributions that should be integrated.

From this perspective, systemic risks of AI appear not primarily as low probability, high impact 'tail risks' by technical errors, malicious attacks or misuse by single actors, or as resulting from specific high impact capabilities of 'frontier AI', even though we do not rule out the possibility that such risks exist. Instead, systemic risks are rooted in continuous social and sociotechnical processes, such as mass application and integration of AI into a multitude of economic, social, and governmental practices as well as business models. Systemic risks are furthermore grounded in coordination and information problems, inappropriate overreliance on AI, dependencies between components, actors and AI systems, domains or sectors, feedback dynamics and rebound effects, governance and institutional frameworks with inappropriate incentives and constraints as well as normative ambiguities. In conclusion, the assessment, prevention, mitigation and governance of systemic AI risks need to be informed by a systemic risk concept that accounts for the continuous aggregations of risks and the interaction of risk phenomena through the lens of complexity, while keeping in mind the core issues of externalities and collective action problems.

## Acknowledgement

The authors would like to thank Wolfgang Eppler for helpful comments as well as Sylke Wintzer and Miriam Micklitz for proof reading and language editing.

## Funding

The work is funded by the German Ministry of Research, Technology and Space (Grant Nr. 01IS23075), which is gratefully acknowledged.

## Declarations

**Use of AI tools.** Parts of the text were first written in German and translated with an AI translation tool (DeepL). The outcomes were reviewed and edited by a person.

**Conflict of interest.** The authors declare no competing interests.